\pdfoutput=1
\documentclass[8.5pt,twoside,twocolumn]{article}
\usepackage[utf8]{inputenc}
\usepackage[super,sort&compress,comma]{natbib} 
\usepackage{mhchem}
\usepackage{times,mathptmx}
\usepackage{sectsty}
\usepackage{balance} 

\usepackage{graphicx} 
\usepackage{lastpage}
\usepackage[format=plain,justification=raggedright,singlelinecheck=false,font=small,labelfont=bf,labelsep=space]{caption} 
\usepackage{fancyhdr}
\begin{document}

\thispagestyle{plain}
\fancypagestyle{plain}{
\renewcommand{\headrulewidth}{1pt}}
\renewcommand{\thefootnote}{\fnsymbol{footnote}}
\renewcommand\footnoterule{\vspace*{1pt}%
\hrule width 3.4in height 0.4pt \vspace*{5pt}} 
\setcounter{secnumdepth}{5}

\makeatletter 
\def\subsubsection{\@startsection{subsubsection}{3}{10pt}{-1.25ex plus -1ex minus -.1ex}{0ex plus 0ex}{\normalsize\bf}} 
\def\paragraph{\@startsection{paragraph}{4}{10pt}{-1.25ex plus -1ex minus -.1ex}{0ex plus 0ex}{\normalsize\textit}} 
\renewcommand\@biblabel[1]{#1}            
\renewcommand\@makefntext[1]%
{\noindent\makebox[0pt][r]{\@thefnmark\,}#1}
\makeatother 
\renewcommand{\figurename}{\small{Fig.}~}
\sectionfont{\large}
\subsectionfont{\normalsize} 

\fancyfoot{}
\fancyfoot[RO]{\footnotesize{\sffamily{1--\pageref{LastPage} ~\textbar  \hspace{2pt}\thepage}}}
\fancyfoot[LE]{\footnotesize{\sffamily{\thepage~\textbar\hspace{3.45cm} 1--\pageref{LastPage}}}}
\fancyhead{}
\renewcommand{\headrulewidth}{1pt} 
\renewcommand{\footrulewidth}{1pt}
\setlength{\arrayrulewidth}{1pt}
\setlength{\columnsep}{6.5mm}
\setlength\bibsep{1pt}

\twocolumn[
  \begin{@twocolumnfalse}
\noindent\LARGE{\textbf{Dynamics and stability of icosahedral Fe-Pt nanoparticles
}}
\vspace{0.6cm}

\noindent\large{\textbf{Paweł T. Jochym,$^{\ast}$\textit{$^{a}$} 
Jan Łażewski,\textit{$^{a}$}
Małgorzata Sternik\textit{$^{a}$} and
Przemysław Piekarz\textit{$^{a}$} }}\vspace{0.5cm}


\vspace{0.6cm}

%
%

\noindent \normalsize{
The structure, dynamics and stability of Fe-Pt nanoparticles have been investigated using DFT-based techniques: total energy calculations and DFT molecular dynamics.
The investigated systems included multi-shell and disordered nanoparticles of iron and platinum. 
The study is concerned with icosahedral particles with magic number of atoms (55): iron-terminated \ce{Fe43Pt12}, platinum-terminated \ce{Fe12Pt43}, and disordered \ce{Fe27Pt28}. 
Additionally, the \ce{Fe6Pt7} cluster has been investigated to probe behaviour of extremely small Fe-Pt particles. 
Molecular dynamics simulations have been performed for a few temperatures between $T=150-1000$ K.
The calculations revealed high structural instability of the Fe-terminated nanoparticles and a strong stabilising effect of the Pt-termination in the shell-type icosahedral particles. 
The platinum termination prevented disordering of the particle even at $T=1000$ K indicating very high melting temperatures of these Fe-Pt icosahedral structures. 
The analysis of evolution of the radial distribution function has shown significant tendency of Pt atoms to move to the outside layer of the particles -- even in the platinum deficient cases.}
\vspace{0.5cm}
 \end{@twocolumnfalse}
  ]

\footnotetext{\textit{$^{a}$~Department of Computational Material Science, Institute of Nuclear Physics PAN, Cracow, Poland ; E-mail: pawel.jochym@ifj.edu.pl}}


\section{\label{sec:intro}Introduction}

Nanoalloys are bi- or multi-component metallic nanoparticles (NPs), with the complex structures and properties, which depend on their size, structure, and composition.\cite{ferrando2008, johnston2008, ortigoza2008, mariscal2012, duan2013}
The size of nanoalloys (1--100 nm) places them between small molecules and bulk crystals.
They can be considered as the intermediate phase sharing common properties of atoms and solid materials. By adjusting the size, geometry and chemical composition, nanoalloys can be optimised for various applications in catalysis,\cite{campbell2004} nanomedicine,\cite{huang2007} optics,\cite{iskandar2009} data recording\cite{sun2000} and energy storage.\cite{duan2014}

Iron-platinum (Fe-Pt) nanoalloys consist of two metallic components with significant ratios (Pt/Fe) of masses (195.1 / 55.8 in amu), bulk moduli (230 / 170 in GPa) and atomic radii (1.46 / 1.30 in \AA) which result in very interesting, not only geometrical, features. 
They belong to the most studied systems due to an extremely high uniaxial magnetocrystalline anisotropy ($K_u=7\times 10^6$ J/m$^3$), which makes Fe-Pt a promising material for applications in ultra-dense magnetic recording media.\cite{sun2000} 
As evidenced by many studies, Fe-Pt systems exhibit enhanced catalytical properties, comparing to pure platinum surfaces.
Fe-Pt nanoalloys may be used as a powerful catalyst for the electro-oxidation of formic acid in fuel cells \cite{chen2006,kim2010,guo2012} and as highly efficient oxidation catalysts for degradation of organic pollutants.\cite{hsieh2012}
Fe-Pt NPs are also promising candidates for biomedical {\it in vivo} applications (e.g. for cellular magnetic resonance imaging\cite{chen2010}) because of their high stability in the presence of oxygen.\cite{kim2005} 
By coating with inorganic molecules, the core-shell Fe-Pt nanoparticles become optically detectable \cite{seemann2014} and  can be functionalised for the hyperthermia treatment of cancer and advanced radiopharmaceutical applications.\cite{seemann2015} 

Due to small sizes, standard diffraction techniques cannot be applied for NPs and most information about their structural properties comes from high-resolution transmission electron microscopy.\cite{tan2006,wang2008,wang2009} 
Relevant information about structural stability and electronic properties of the Fe-Pt nanoclusters have been obtained by the theoretical studies based on either semiempirical potentials or {\it ab initio} techniques.\cite{fortunelli1999,chepulskii2005a,chepulskii2005b,muller2005,yang2006,gruner2008} 
The Monte Carlo simulations showed that a continuous transformation between the ordered structure $L1_0$ ($tP4$) and a disordered phase in Fe-Pt NPs occures at temperatures lower than the bulk melting temperature (1572~K).\cite{chepulskii2005a,chepulskii2005b} 
Disordering processes in nanoalloys are enhanced due to finite-size and surface effects, e.g. strong surface segregation tendency of Pt atoms.\cite{muller2005,yang2006} 
The structural order and magnetic properties of Fe-Pt nanoalloys with various morphologies and chemical composition have been studied previously within the large-scale Density Functional Theory (DFT) calculations.\cite{gruner2008} 
In the present study, we focus on icosahedral Fe-Pt NPs, which are energetically favoured over the $L1_0$ phase for small particles.\cite{tan2006,wang2008,gruner2008}
Application of DFT-based methods, including DFT molecular dynamics, provides additional, independent information on the behaviour of Fe-Pt NPs, which supplement the published studies based mainly on semiempirical potentials and supplies a valuable consistency check on the already established models of these systems. 

Our study of Fe-Pt NPs has been divided into two main parts.
The first part is a set of total energy calculations and structural optimisations, designed to establish static stability properties of the investigated systems. 
The methods employed in this part are described in section \ref{ssec:structopt} and the results in section \ref{ssec:static}.
The second part of the work concerns the investigation of dynamical stability of the systems using DFT-based molecular dynamics (MD). 
Description of the methods employed for this part of the work is included in section \ref{ssec:mdcalc} and the results are discussed in section \ref{ssec:md}.
Section \ref{sec:rnd} contains the results of the investigation of dynamical stability of the disordered \ce{Fe6Pt7} and \ce{Fe27Pt28} particles. 
The concluding section \ref{sec:concl} summarizes the key results of the paper.

\section{\label{sec:method}Methods}

A common denominator of the methods used in the present work is the Density Functional Theory.\cite{hohenberg1964,kohn1965}
Both parts of our analysis use the same DFT code as implemented in Vienna {\it Ab initio} Simulation Package (VASP)\cite{kresse1996a,kresse1996b} and the same atomic data sets provided with this package.
For the preparation and analysis of the molecular dynamics runs and control of the DFT computations, we have used standard, python-based software stack\cite{IPython,ASE,SciPyLib,NumPy,MPL}.
The details of the computational setups used for the work are described below -- separately for each part of the analysis.

\subsection{\label{ssec:structopt}Structural optimisation}

The static, zero Kelvin temperature calculations were performed using the full-potential projector-augmented wave method\cite{blochl1994,kresse1999} within the GGA approach in PAW-PBE form.\cite{perdew1992,perdew1996} 
The following valence base configurations were included: Fe $3d^74s^1$ and Pt $6s^15d^9$. 
The integrations in the reciprocal space was reduced to the single {\bf k}-point $\Gamma$ and the energy cut-off for the plane waves expansion was equal to 320 eV. 
The crystal structure was optimised using the conjugate gradient technique with the energy convergence criteria set at $10^{-7}$ and $10^{-5}$ eV for the electronic and ionic iterations, respectively. 
In order to prevent interactions between system images created due to the periodic boundary conditions used in VASP, each nanoparticle was placed in the big 20\AA$\times$20\AA$\times$20\AA{} box containing about 10 \AA{} wide vacuum. 
This limited force constants between neighbouring images below 0.3\% of the strongest force constant in the system.

\subsection{\label{ssec:mdcalc}Molecular dynamics}

The molecular dynamics simulations were carried out in the same DFT framework as the static calculations described above and with the set of methods and tools used in our previous work on minerals.\cite{jochym2010}
We used the PAW-PBE atomic data sets and the spin-polarised VASP calculation for the DFT part.
To lower the computational cost we adjusted the accuracy parameters of the calculation -- since the MD does not need extremely precise inter-atomic forces.
The precision parameter of the calculation was set to Normal, the cut-off energy was reduced to 275 eV and the energetic convergence threshold was set at $10^{-6}$~eV.
A small numerical noise added by lowering accuracy of the calculation is drowned by the thermal noise of the system. 
Furthermore, additional energy pumped into the system by this noise is cancelled out by the action of the thermostat.
The actual molecular dynamics used the VASP implementation of the Nose thermostat MD algorithm with 10~fs time step, which was determined by convergence testing.
The use of this fairly large time step was possible thanks to large masses of atoms in the system.
The starting configurations correspond to perfect icosahedral geometries with the multi-shell or disordered structures.
In case of the disordered structure the atomic sites were assigned randomly to one of the species by drawing from the pre-determined set of atoms -- 6 Fe and 7 Pt or 27 Fe and 28 Pt, respectively.
The resulting structures were pre-optimised, by standard static methods (see sec.~\ref{ssec:structopt}) to avoid large inter-atomic forces at the start of the MD run.
For the same reasons we have used the results of the static structural optimisation (sec. \ref{ssec:static}) for the starting configuration of ordered particles. 

\begin{figure}[h]
\centering
  \includegraphics[width=\columnwidth]{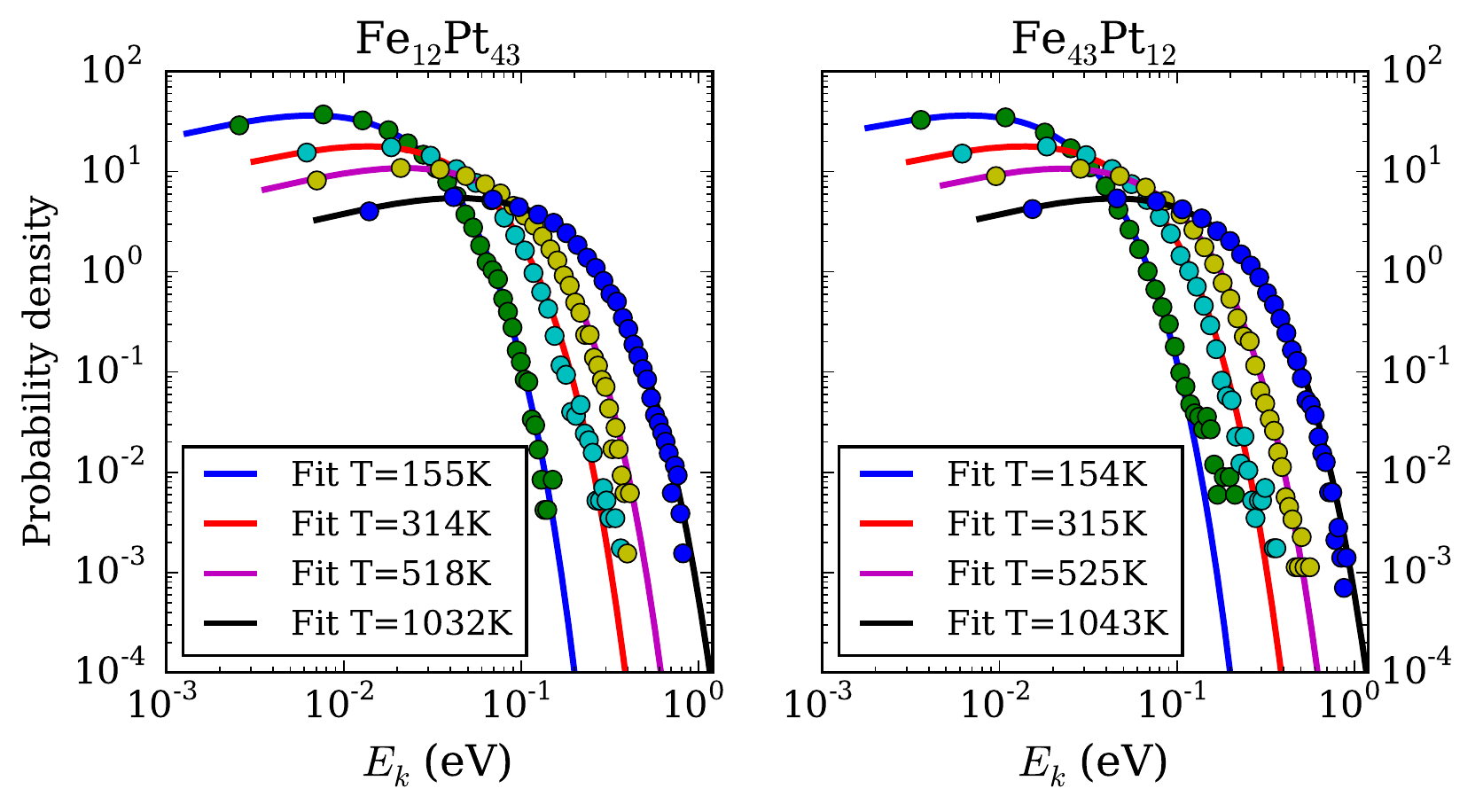}
  \caption{Distribution of particle kinetic energy in the icosahedral-shell system (dots) and Maxwell-Boltzmann distribution fits for each data set (lines). The best-fit temperature for each data set is included in the legend.}
  \label{fig:Ek-dist}
\end{figure}

\begin{figure}[h]
\centering
  \includegraphics[width=\columnwidth]{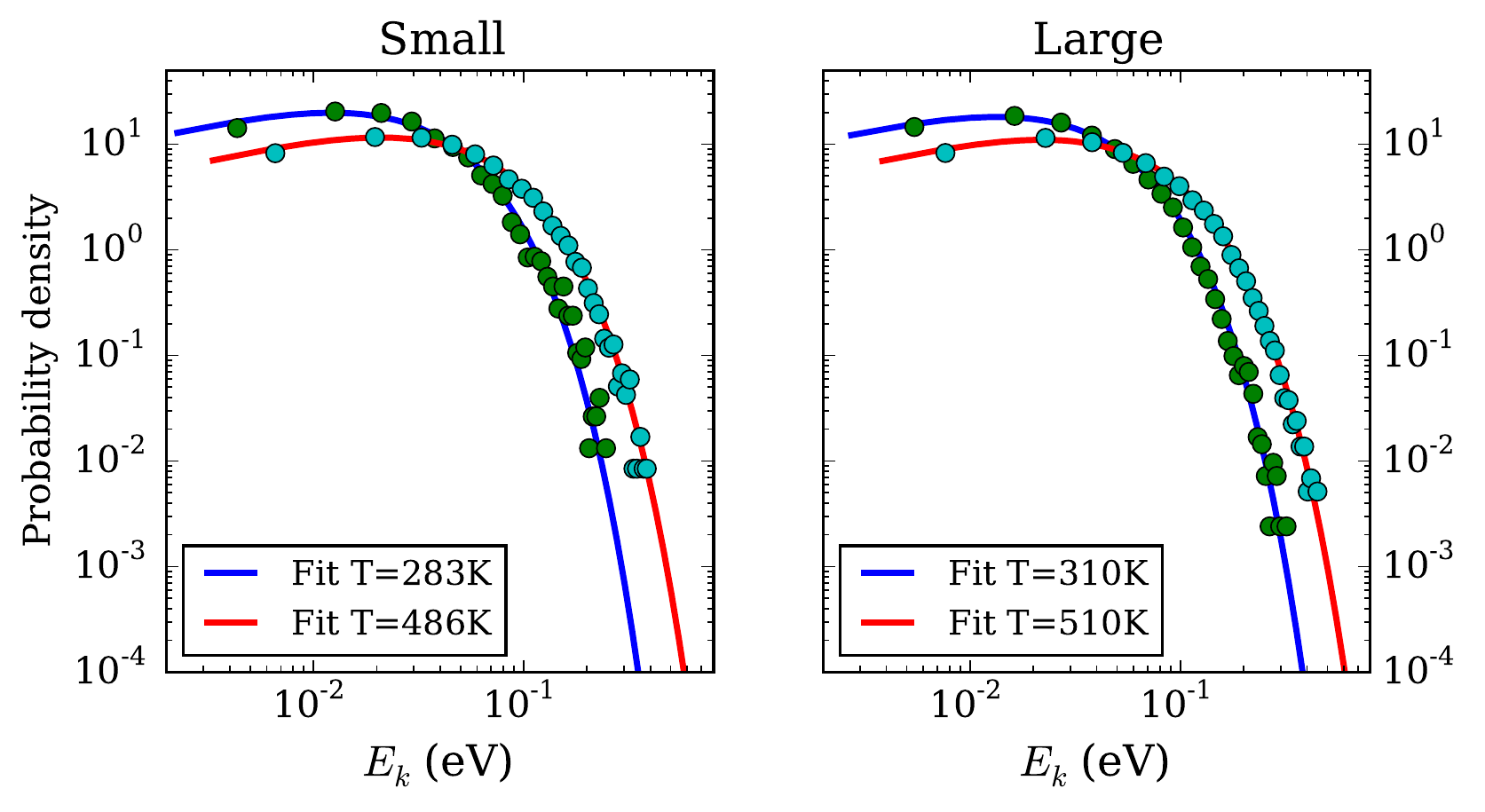}
  \caption{Distribution of particle kinetic energy in the random systems (dots) and Maxwell-Boltzmann distribution fits for each data set (lines). 
'Small' system (left column) is a disordered \ce{Fe6Pt7} cluster, while 'Large' system (right column) is a disordered \ce{Fe27Pt28} cluster (see sec. \ref{sec:rnd}).
The best-fit temperature for each data set is included in the legend.}
  \label{fig:Ek-rnd}
\end{figure}

The calculation procedure involved a thermalisation run of the system with kinetic energies assigned to atoms according to the Maxwell distribution.
To assess the level of thermalisation of the system we calculated a distribution of kinetic energies in each step for all the atoms and for an ensemble configurations in the 0.1 ps sliding window (10 steps).
The average energies in each step were compared with the value expected for the perfect Maxwell distribution.
When the systematic drift of the computed average value transitioned into random-looking fluctuations, we performed an additional check for good thermalisation of the system by computing the full distribution of kinetic energies in the ensemble built from the configurations in the 1 ps window (100 steps) and comparing it with the Maxwell distribution.
The comparison was performed by fitting the Maxwell distribution to the histogram of the kinetic energy in the ensemble and comparing the obtained temperature with the expected value.
We also checked if the size of the residuals is consistent with the size of the fluctuations expected for the ensemble.

The combination of all such comparisons is presented in Figs \ref{fig:Ek-dist} and \ref{fig:Ek-rnd}.
The probability distributions presented in these figures are based on the whole measurement part of the MD run -- the thermalisation part was discarded.
Very good agreement of data points with the perfect Maxwell distribution is slightly broken only at the extreme large energy range in the case of the Fe-terminated system at the lowest temperature ($T=150$ K).
Notice the logarithmic scale on both axis in Figs \ref{fig:Ek-dist} and \ref{fig:Ek-rnd}, which was chosen to amplify any discrepancy from the expected distribution -- on the linear scale the discrepancy is invisible.
This is the range of a high-energy tail of the distribution and one can fully expect large relative fluctuations there, due to the small probability density in this region and thus, small counts in the histogram bins.
Furthermore, as the calculations revealed, the \ce{Fe43Pt12} is a quite unstable system which results in difficulties with bringing it to equilibrium at low temperatures.
Nevertheless, we do not consider this single discrepancy to be significant for the analysis carried out in this work and regard all investigated systems as adequately thermalised.

\section{\label{sec:struct}Icosahedral Fe-Pt nanoparticles}

Icosahedron (Ih) is one of the five Platonic solids and one of the non-crystallographic structures with fivefold symmetries.\cite{mackay1962}
It consists of 20 surfaces, 12 vertices and 30 edges, each surface has close-packed structure with the (111) orientation. 
The most stable particles are built of closed shells with the number of atoms $N_{shell}=1,12,42,\dots$.
Fe-Pt NPs crystallising in the icosahedral geometry with the alternating Fe and Pt shells exhibit very high structural stability.\cite{wang2009}
The stability of Ih NPs results from the low energy of the close-packed (111) surfaces, which compensates the internal stress existing in the particle core.   
The present study focuses on two types of icosahedral particles: particles with perfect shell structures (terminated with Pt or Fe atoms) and disordered NPs with approximately equal number of Fe and Pt atoms.
The ordered \ce{Fe12Pt43} particle with the alternating shell structure: 1 Pt, 12 Fe and 42 Pt atoms is displayed in Fig.~\ref{fig:pfp150}(a).  

\subsection{\label{ssec:static}Static calculations}

\begin{figure}[h]
\centering
  \includegraphics[width=0.95\columnwidth]{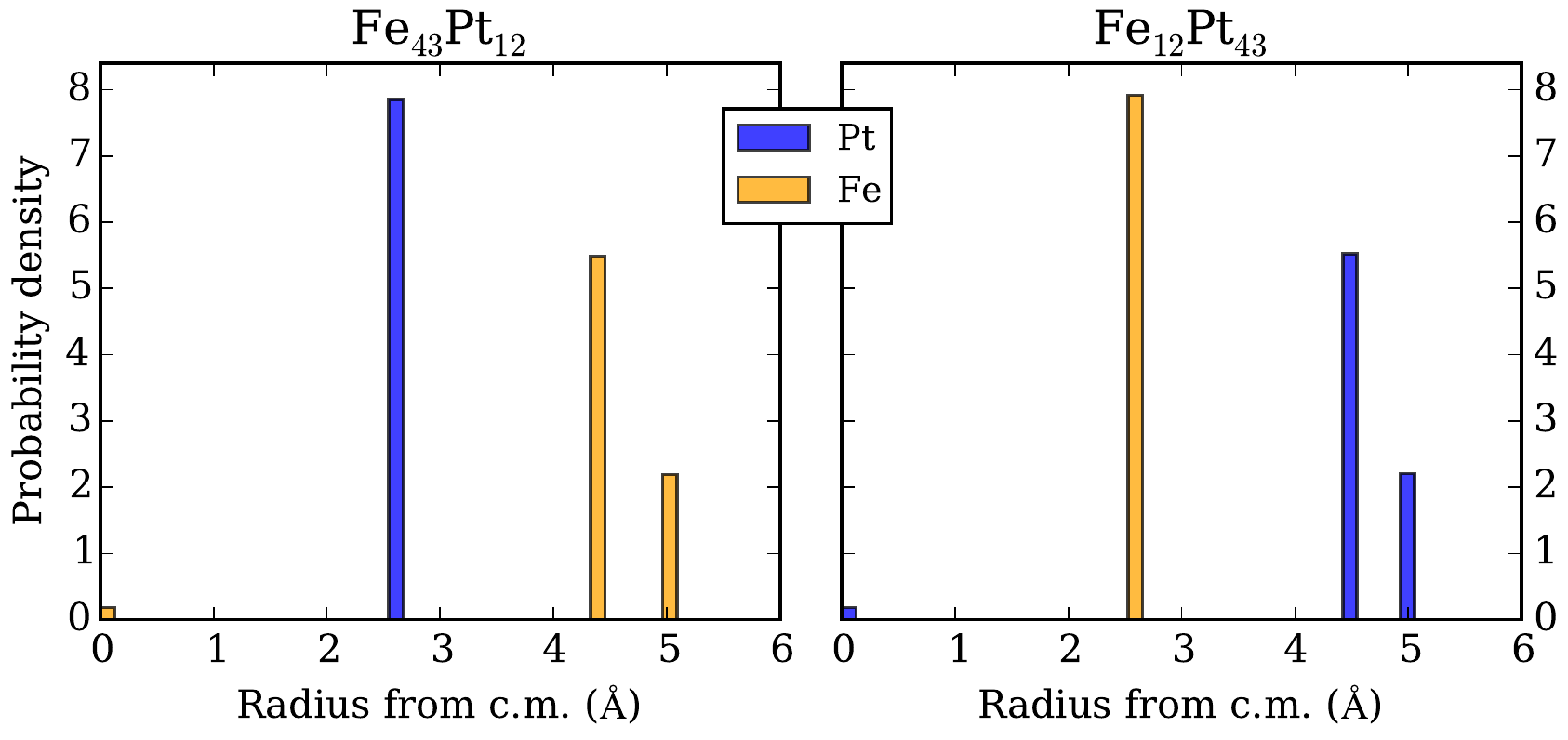}
  \caption{Radial distribution of atomic positions from the mass centre of \ce{Fe43Pt12} (left panel) and \ce{Fe12Pt43} (right panel).}
  \label{fig:rdf-static}
\end{figure}

The first step, after constructing individual NPs, was to optimise the structure at $T=0$~K, by minimising the total energy of the system, keeping $T_h$ point group symmetry elements only, since $I_h$ symmetry was broken in periodic boundary conditions.\footnote{$I_h$ point group includes 5-fold axes, which are not compatible with the translational symmetry and 3D periodic boundary conditions imposed on supercell in VASP package. Remaining elements of $I_h$ group, forming $T_h$ point group, were considered in our static calculations.}
The lowering of particle symmetry allows not only for volume changing but also for surface shell relaxation. 
In both cases, \ce{Fe43Pt12} and \ce{Fe12Pt43}, high symmetry configuration with well-defined alternating Fe/Pt shells was obtained. 
The calculated radial distribution functions of optimised structures are presented in Fig.~\ref{fig:rdf-static}. 
In icosahedra there is only one non-equivalent position of atoms in the first shell, which defines the distance between the centre and the first shell, and two different positions in the second shell: at vertices and edges of the icosahedron.  
Interestingly, in both NPs, the distances from the central atom to the first shell (2.643 \AA{} and 2.587 \AA{} for \ce{Fe43Pt12} and \ce{Fe12Pt43}, respectively) as well as to vertices in the second shell (5.091 \AA{} for \ce{Fe43Pt12} and 5.054 \AA{} in \ce{Fe12Pt43} case) are quite similar.
A larger difference ($\sim0.1$ \AA) is found for the distance between the centre and edge atoms in the second shell (4.374 \AA{} in \ce{Fe43Pt12} and 4.477 \AA{} in \ce{Fe12Pt43} case).

In the surface shell, there are two characteristic nearest neighbours (NN) distances: from a vertical atom to the nearest edge atoms (1--2 in Fig.~\ref{fig:pfp150}) and between two nearest edge atoms (2--2 in Fig.~\ref{fig:pfp150}). 
In \ce{Fe43Pt12}, they read 2.704~\AA{} and 2.677~\AA{}, while in \ce{Fe12Pt43}  -- 2.767~\AA{} and 2.663~\AA{}, respectively. 
In the Pt subsurface shell NN distances are equal to 2.778 \AA, while in the Fe one they are 2.720 \AA.
For comparison, in the {\it bcc} iron structure simulated with the same atomic data set,\cite{lazewski2006} the NN distance is equal to 2.485~\AA{} and in {\it fcc} platinum 2.772~\AA. 
This means that, contrary to platinum shells, the inter-atomic distances in both iron shells, the surface one of \ce{Fe43Pt12} and subsurface of \ce{Fe12Pt43}, are strongly stretched ($\sim8-10\%$) which can be one of the main reasons of the iron-terminated NP instability.
The similar influence of strain on iron surface stability was observed previously in the Fe monolayer on W(110) surface\cite{lazewski2007} and in FeAu and FePt multilayers\cite{sternik2006,jochym2008}.
In both particles, the NN distances between the Fe and Pt atoms located in the first and second shell ($\sim2.50$ \AA) and between the central atom and the first shell ($\sim2.60$ \AA) are reduced comparing to the Fe-Pt distance in the bulk {\it fct} $L1_0$ structure (2.70 \AA).\cite{couet2010} 
It causes the internal stress in the core, which may be the driving force for the amorphisation processes and may induce the structural transformation observed in larger Fe-Pt particles.

\subsection{\label{ssec:md}Molecular dynamics}

The dynamical stability of a nanoparticle is quite different from the static one. 
The later requires only that the system is in the \emph{local} energy minimum, the former necessitates stability of the structure against finite, even large, thermal fluctuations.
The amplitude of fluctuations is determined by the temperature of the system and the strength of inter-atomic bonds present in the system.
To go beyond the static stability analysis described above, we need to include temperature in our calculations. 
We used the DFT-MD technique described above (see section~\ref{ssec:mdcalc}). 
The simple way to asses the effects of such calculation is by inspection of the evolution of the structure. 

\begin{figure}[h]
\centering
  \includegraphics[height=3cm]{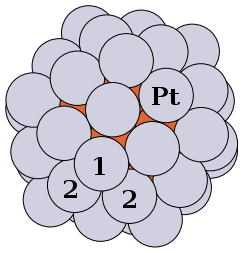}${(a)}$
  \includegraphics[height=3cm]{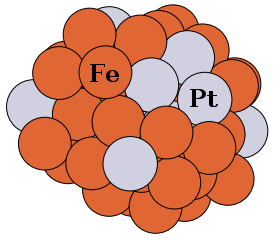}${(b)}$
  \caption{\ce{Fe12Pt43} (a) and \ce{Fe43Pt12} (b) structures at $T=150$ K after 10 ps thermalisation and 20 ps of MD run. The difference in structural stability is visible even at such a low temperature. Elements: Fe -- dark/brown, Pt -- light/grey. 1 and 2 denote two non-equivalent positions of atoms in the surface shell of ideal icosahedron.}
  \label{fig:pfp150}
\end{figure}

\begin{figure}[h]
\centering
  \includegraphics[width=0.95\columnwidth]{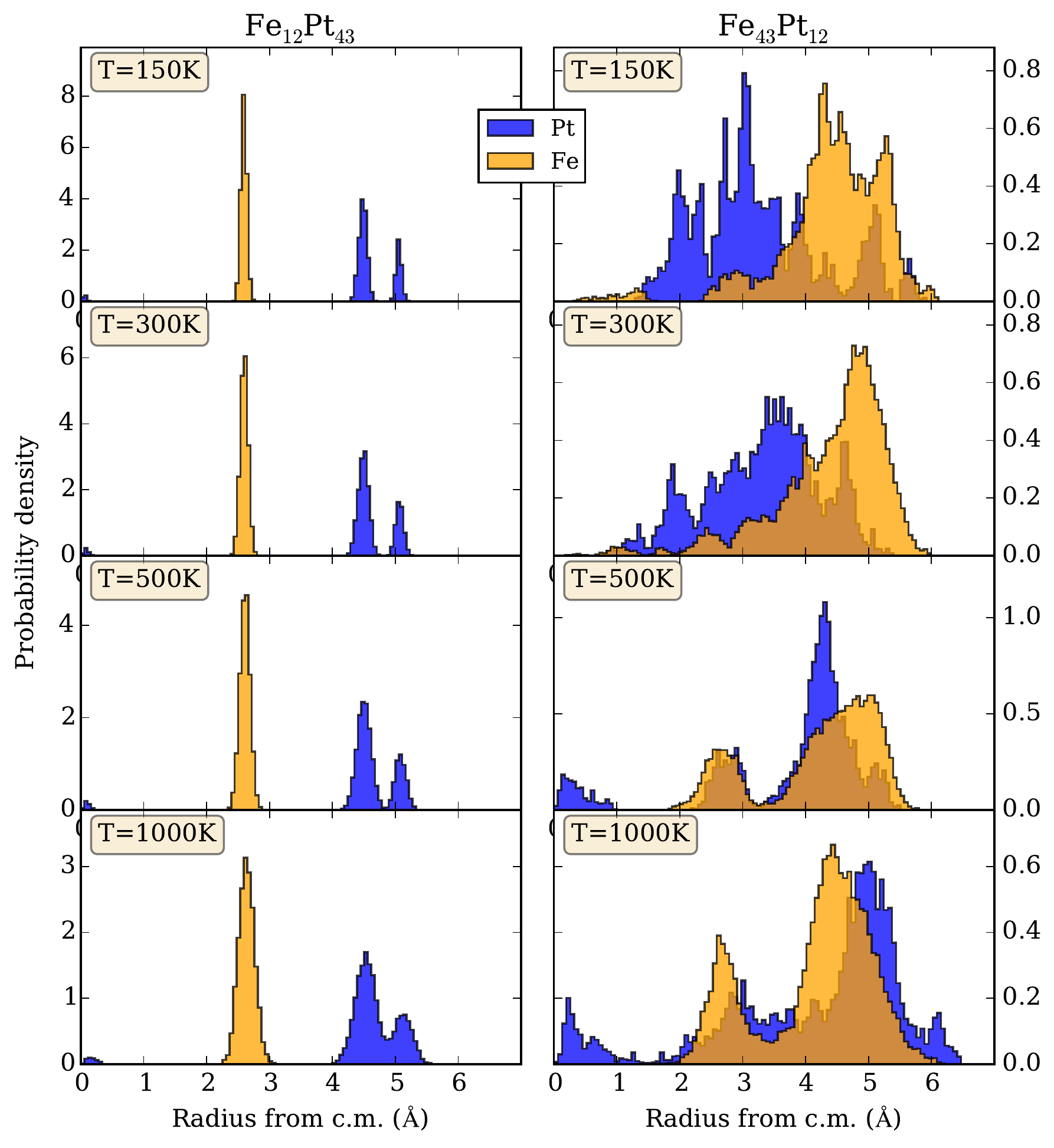}
  \caption{Radial distribution of atomic positions from the mass centre (c.m.) of the icosahedral-shell system. 
The results for all temperatures are compared side-by-side for two investigated systems Pt-terminated (left) and Fe-terminated (right). 
The probabilities are normalised to unity per each element.
The semitransparent colour of the Fe plot (orange) is used to show the Pt plot (blue) in the same range.}
  \label{fig:rdf-order}
\end{figure}

The structures after thermalisation (10 ps) and equilibration (20 ps) runs at target temperature $T=150$ K are depicted in Fig.~\ref{fig:pfp150}.
At the first glance the qualitative difference between platinum-terminated (a) and iron-terminated (b) particles is obvious.
All the Fe atoms are in the subsurface in the former, nearly perfect structure, while there are many platinum atoms on the surface of the later, strongly distorted structure -- despite the fact that this is a Pt-deficient system.
What is more, the picture is virtually identical at higher temperatures.
Even at $T=1000$ K there is no \emph{qualitative} visual difference between Fig.~\ref{fig:pfp150} and the corresponding picture at that temperature.
This calls for more detailed, quantitative analysis. 
In the crystalline material the radial distribution function (RDF) would be a first-line tool for such investigation.
Here, the system is finite and we cannot use standard RDF directly.
Instead we have used radial probability distribution $p_i(r)$ of finding atom of element $i$ at the distance $r$ from the centre of mass of the system.
The probability is calculated over the ensemble of last 10 ps (1000 steps) of configurations at the end of the MD run.
We performed the same calculation on the much smaller ensembles ($0.1-1$ ps) and determined that there is no significant difference except for much larger fluctuations present in the histograms (for obvious statistical reasons).

The probability density $p_i(r)$ plots, normalized to unity for each atomic species, for all investigated icosahedral systems and temperatures are collected in Fig.~\ref{fig:rdf-order}. 
It should be remembered that the systems are started in the perfectly ordered state with temperature-determined kinetic energy distribution.
At the start of the simulation $p_i(r)$ function for all systems resembles the plot for \ce{Fe12Pt43} at $T=150$ K (top, left corner of Fig.~\ref{fig:rdf-order}) or the static RDF from Fig.~\ref{fig:rdf-static}.
The probability distributions in the Fig.~\ref{fig:rdf-order} are calculated at the end of the equilibration.
Since the relaxation processes may have very different time scales, one cannot be sure that the \emph{true} equilibrium state has been reached.
Nevertheless, the lack of systematic drift in the probability distributions at the end of the simulation period allows us to conclude that the systems are reasonably close to equilibrium.

There is a clear difference between final states of the Pt-terminated (left column in Fig.~\ref{fig:rdf-order}) and Fe-terminated (right column in Fig.~\ref{fig:rdf-order}) system.
The former, essentially, exhibits temperature-caused broadening of the distribution peaks, while the later shows clear signs of substantial reconstruction of the structure as well as much stronger broadening and distortion of the peaks.

Since there is not much happening in the Pt-terminated structure, let us concentrate on the Fe-terminated one.
We can distinguish two main phenomena in these plots.
The first one is a significant disordering of the system, shown in very broad and irregular peaks -- even at low temperatures.
The second phenomenon is a systematic drift of the platinum distribution to the external shell of the particle.
We expect the former process to have fairly short time scale and low activation barriers, opposite to the other one which requires global, collective movements and thus large activation energy, and exhibits much longer time scale.
Consequently, the reconstruction is probably close to finished only in the high temperature case ($T=1000$ K bottom of Fig.~\ref{fig:rdf-order}), is under way for intermediate temperatures ($T=500$ K), and is only starting for lower temperatures ($T=150$ K, $T=300$ K). 
Nevertheless, even at $T=150$ K we can see non-vanishing Pt probability at the outside shell and some migration of Fe atoms towards the interior of the particle.
This process seems fully consistent with our earlier observations drawn from the structures in Fig.~\ref{fig:pfp150}.
The high temperature in the $T=1000$ K case helps to overcome the activation barriers and speeds up the reconstruction.
We expect that if we leave the lower temperature cases to evolve long enough they will reach similar shape of the distribution as the high temperature case.
Unfortunately, this kind of time scale is prohibitively long to achieve -- we estimate it will require at least one year worth of MD run for $T=150$~K.
Note also that after reconstruction some level of ordering is recovered in the high temperature system -- the peaks are better defined and localised and actually narrower than for the low-temperature systems.
In fact, one can recognise a clear gradient of decreasing noise and disorder of the distributions with increasing temperature in the right column of Fig.~\ref{fig:rdf-order}.
Interestingly, Fe atoms can be found close to the centre only at low temperatures, while they are systematically replaced by platinum atoms at higher temperatures.

The results obtained for two types of particles can be easily understood if we take into account strong tendency of Pt atoms for occupying the outer layers of NPs.
It prevents disordering of Pt-terminated particles, which keep their perfect multi-shell structure even at high temperatures.
It agrees with the previous theoretical studies\cite{fortunelli1999,chepulskii2005a,chepulskii2005b,muller2005,yang2006,gruner2008} and the experimental observations of good thermal stability of icosahedral Fe-Pt particles even with much larger sizes.\cite{wang2008} 
In the thin films, the Pt-termination was deduced from the DFT calculations of the phonon spectra and the surface sensitive nuclear inelastic scattering
measurement.\cite{couet2010} The simulated Fe-terminated surface showed strong deviation from the bulk behaviour. 

The crucial role in the stabilization of multi-twinned nanoalloys is played by the difference in atomic sizes of the two elements: larger size mismatch reduces the compression of the core. 
The MD simulations performed on the polyicosahedral core-shell Ag-Cu and Ag-Ni clusters revealed much higher melting temperatures than for pure Ag, Cu and Ni particles.\cite{rossi2004,ferrando2005} 
Even single impurities can stabilize the icosahedral Ag particles -- the smaller the doping atom, the higher the melting temperature.\cite{mottet2005}
As shown in recent studies, the stability of Ni-Au NPs is also enhanced in the core-shell structure, which has much higher melting temperature than random alloys.\cite{li2014}  
     
The \ce{Fe43Pt12} particle is very unstable due to the movement of Pt atoms towards the surface.
This leads to significant changes of atomic positions and strong deformation of the icosahedral geometry.
Apart from the segregation processes resulting from different sizes of constituent atoms, other amorphisation mechanisms may also be of some importance here.
The previous MD simulations revealed the amorphisation of mono-atomic \ce{Pt55} icosahedral particles involving the rosette-like structural transformation, which forms a sixfold ring centred around the fivefold vertices and breaks the $I_h$ symmetry.\cite{apra2004}
A preference for low-symmetry amorphous structures was found also in the Au nanoclusters.\cite{garzon1998}

\section{\label{sec:rnd}Disordered Fe-Pt nanoparticles}

The results described above concerned icosahedral multi-shell structures.
These structures are inherently non-stoichiometric and thus quite different from the fragments of the bulk crystals.
On the other hand, such highly ordered small particles are very unlikely to form naturally, although advanced experimental techniques may be well capable of producing just such systems.
The synthesised icosahedral particles with larger sizes ($\sim5-6$ nm) and close to equiatomic compositions exhibit the shell-periodic structure with the Fe/Pt core and the Pt enriched outer shell.\cite{wang2008} 
For small systems such ordered core-shell structures may be very unstable and we expect natural processes to form rather disordered particles with the composition of elements determined by their relative concentrations in the environment and other factors such as chemical potentials, surface energies, bonding energies etc.

\begin{figure}[h]
\centering
  \includegraphics[height=1.5cm]{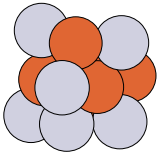}${(a)}$
  \includegraphics[height=3cm]{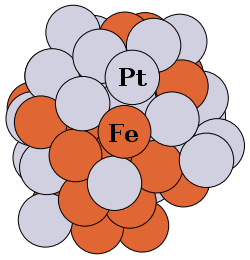}${(b)}$
  \includegraphics[height=3cm]{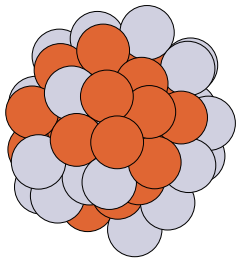}${(c)}$
  \caption{Small disordered particle at $T=300$ K (a), larger disordered particle at $T=300$ K (b) and at $T=500$ K (c). All structures after 30 ps thermalisation. Elements: Fe -- dark/brown, Pt -- light/grey.}
  \label{fig:rndstruct}
\end{figure}

As the simplest model we have decided to select two \emph{magic number} (i.e. numbers from the partial sum: $1+12+42+\dots$) particles with concentrations close to 50/50 (the magic numbers are odd, thus the exact 50\% concentration is unattainable).
The selected structures are \ce{Fe6Pt7} and \ce{Fe27Pt28} -- both with slight iron deficiency -- as we expect such particles to be more stable 
than their platinum-deficient counterparts (see above).
We selected magic-number particles (13 and 55 atoms respectively) to make it possible for the systems to relax to closed-shell configurations which one can expect to be energetically favourable.

The applied procedure was identical to the one used above in section~\ref{ssec:mdcalc}.
The only notable difference was a much longer thermalisation period required for the system to settle down close to the equilibrium -- of the order of 10~ps.
Due to a large size of the configuration space of disordered particle and a long time scale of slow relaxation processes, we do not expect the system to reach the true equilibrium state.
Nevertheless, by examining the directions the system configuration drifts after thermalisation, we can determine the direction of these relaxation processes.
This idea can be further cross-checked by application to very small particles.
Due to much smaller configuration space and much lower barriers for collective movements, a small particle can reach its equilibrium in much shorter time.
Running the MD simulation until full relaxation is actually feasible in this case.
We ran this procedure for the smallest non-trivial, magic cluster: \ce{Fe6Pt7} at $T=300$~K and $T=500$~K.
The parameters of the simulation were the same as for larger systems, except that the total simulation length of 30~ps was easy and inexpensive to achieve.
Simultaneously, the small system stopped to drift in any significant way after only few picoseconds of simulation -- indicating arrival at the equilibrium configuration.
The resulting structures (at the end of the MD run) for both sizes of the particles are depicted in Fig.~\ref{fig:rndstruct}.

\begin{figure}[h]
\centering
  \includegraphics[width=0.95\columnwidth]{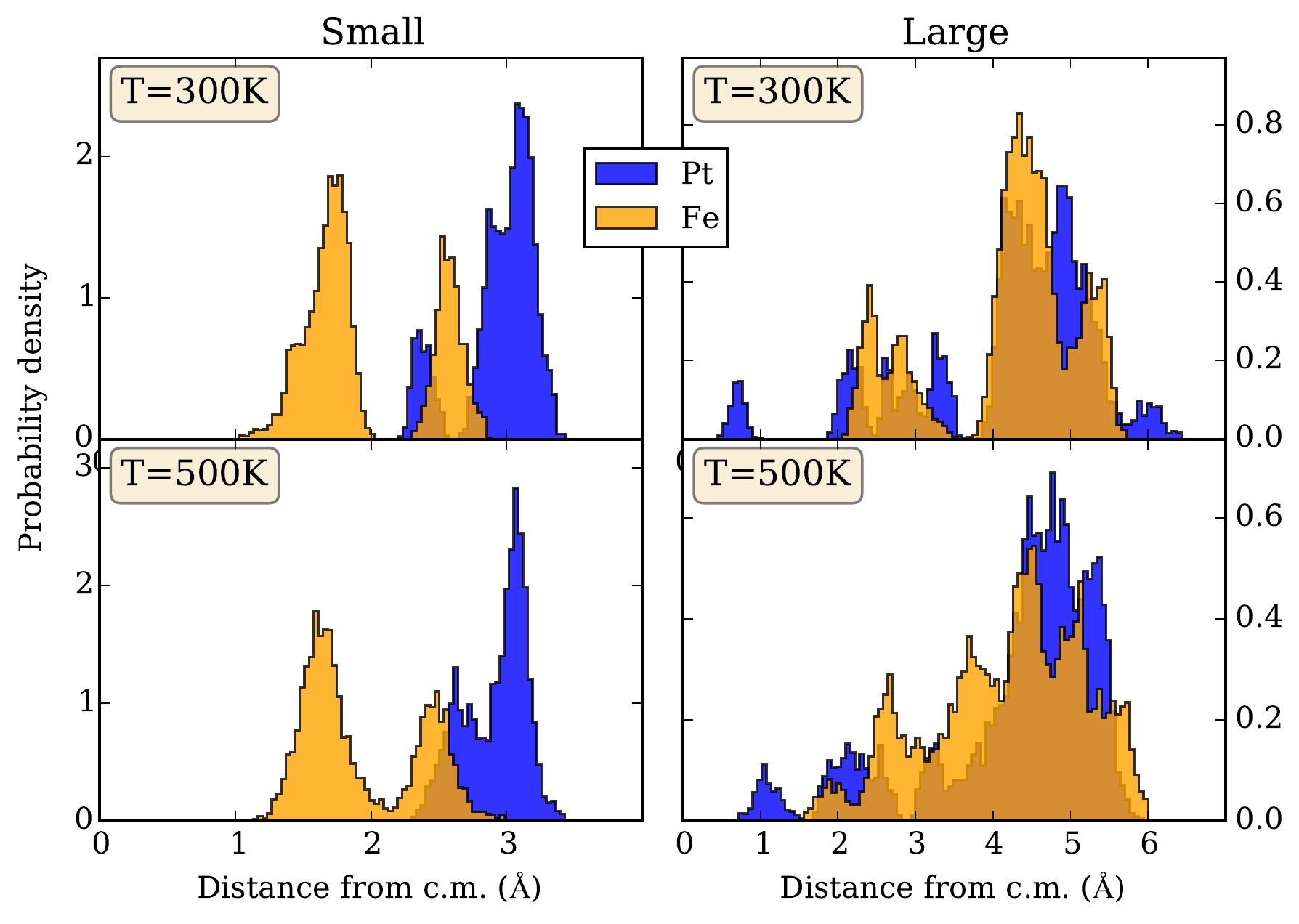}
  \caption{Radial distribution of atomic positions from the mass centre (c.m.) of the system. The results for all temperatures are compared side-by-side for two investigated systems small (left) and large (right).
The semitransparent colour of the Fe plot (orange) is used to show the Pt plot (blue) in the same range.}
  \label{fig:rdf-rnd}
\end{figure}

We applied the same analysis techniques as described above for icosahedral particles.
The resulting radial distribution functions are plotted in Fig.~\ref{fig:rdf-rnd}.
The results for the small system indicate the same pattern we observed in the icosahedral particles -- the platinum atoms tend to move to the external layer of the particle and the iron atoms gather closer to the centre.
The same pattern is less visible in the data for the larger particle -- which is not as close to equilibrium as the small one -- but the same trend is visible anyway.
There is clear difference in the level of ordering exhibited by small versus larger particles, for the reasons explained above, but even for a larger system there are clear signs of the emerging structure in the form of obvious peaks in the RDF function and visible movement of platinum atoms to the outside of the particle.
The details of the RDF distribution are subject to standard short-term statistical and thermal fluctuations, but the overall shape of the distributions proved to be fairy robust and subject only to long time scale relaxation processes.
We believe that if the system is left to evolve even longer, it will probably reach partially ordered equilibrium with platinum atoms located mainly on the surface of the particle.
Unfortunately, the expected time scale -- measured at least in nanoseconds, perhaps even longer -- is prohibitive for the pure DFT MD technique.
This type of calculation would require a method which is at least few orders of magnitude more effective than pure DFT MD -- e.g. neural network derived, multi-parameter effective potentials.\cite{hobday1999,malshe2008,springborg_neural_2010}

\section{\label{sec:concl}Conclusions}

In the presented work we have investigated the stability and dynamics of icosahedral nanoparticles constructed with two radically different (in terms of mass, atomic radius, bulk modulus, magnetic moment) metals: iron and platinum.
The optimised icosahedral NPs with perfect layered structures have rather stiff platinum shells with inter-atomic NN distances well corresponding to the Pt bulk values and strongly stretched iron shells with NN distance elongated by $8-10$\% in comparison to the bulk. 
Even for iron-separated NPs, with significant dominance of Fe atoms, compression of Pt shells does not occur and proves to be impossible.
On the other hand, unstable iron shells drive the system to reconstruction and/or significant lowering of the melting temperature.

The DFT molecular dynamics calculations carried out in a number of temperatures ($T=150-1000$~K) revealed strong instability of the iron-terminated structure and significant tendency for platinum atoms to migrate into surface layer -- even at low temperatures and platinum-deficient particles.
The same tendency was observed in the very small (13 atoms) as well as larger disordered particles (55 atoms) with atomic concentrations close to the 50/50 ratio.
Furthermore, the platinum capping layer seems to additionally stabilise the particle.
In our calculations the platinum-terminated system proved to be stable up to $T=1000$~K.
On the other hand, the iron-terminated system shows substantial instability -- leading to significant reconstruction of the structure and probably partial melting even at a low temperature ($T=150$~K).

The stabilizing effect of platinum atoms seem to be related to the geometric aspect of platinum enclosure of the NP. 
In the case of \ce{Fe43Pt12} NP, platinum atoms have no significant influence on the stability of the NP because they only decorate the surface instead of forming complete enclosure.
The case of 50/50 Fe-Pt NPs further indicates that it is rather geometry than concentration of platinum atoms that effects the stability of nanoparticles.

The structural properties of small Fe-Pt NPs and their dependence on temperature discussed in the present work may be important for nano-technological applications.  
For instance, the catalytic activity of strongly distorted NPs could be enhanced comparing to pure Pt surfaces or larger Fe-Pt particles. 
A larger distribution of Fe-Pt bond lengths as well as the presence of both types of atoms at the surface may lead to selective modification of electronic structure and catalytic activity of Pt atoms.
It was observed that additional Co atoms decorating the surface of Pt nanocrystals lead to increased selectivity of hydrogenation reactions.\cite{tsang2008} 

\section{\label{sec:ack}Acknowledgements}

This work was partially supported by the COST Action MP0903 "Nanoalloys as Advanced Materials: From Structure to Properties and Applications" and by the Polish National Science Centre (NCN) under Project No. 2011/01/M/ST3/00738.

\footnotesize{
\bibliography{FePt} 
\bibliographystyle{rsc} 
}

\end{document}